\documentclass[10pt,
               a4paper,
               rmp,
               showpacs,
               showkeys,
               twocolumn,
              ]{revtex4}
\usepackage{graphicx}
\usepackage{amsmath,amssymb}
\newcommand{\spliter}
{\vskip1ex
 \centerline{\rule{0.6\columnwidth}{0.3mm}}
 \vskip1ex}

\begin{document}
\title{Recurrence Time Statistics for Finite Size Intervals} 
\author{Eduardo G. Altmann}
\email[Author to whom correspondence should be sent. E-mail address:
      ]{altmann@if.usp.br}
\affiliation{Instituto de F{\'\i}sica, Universidade de S{\~a}o Paulo,
             C.P. 66318, 05315-970, S{\~a}o Paulo, S{\~a}o Paulo,             
             Brazil.}   
\author{Elton C. da Silva}
\affiliation{Instituto de F{\'\i}sica, Universidade de S{\~a}o Paulo,
             C.P. 66318, 05315-970, S{\~a}o Paulo, S{\~a}o Paulo,
             Brazil.}

\author{Iber{\^e} L. Caldas}
\affiliation{Instituto de F{\'\i}sica, Universidade de S{\~a}o Paulo,
             C.P. 66318, 05315-970, S{\~a}o Paulo, S{\~a}o Paulo,
             Brazil.}
\date{\today}

\begin{abstract}
  We investigate the statistics of recurrences  to finite size intervals for
  chaotic dynamical systems.  We find that the typical distribution presents an
  exponential decay for almost all recurrence times except for a few short times
  affected by a kind of memory effect.  We interpret this effect  as being
  related to the unstable periodic orbits inside the interval. Although it is
  restricted to a few short times it changes the whole distribution of
  recurrences.  We show that for systems with strong mixing properties the
  exponential decay converges to the Poissonian statistics when the width of the
  interval goes to zero. However, we alert that special attention to the size of the
  interval is required in order to guarantee that the short time memory effect is
  negligible when one is interested in numerically or experimentally calculated
  Poincar{\'e} recurrence time statistics.  
\end{abstract}

\maketitle

{\bfseries \noindent
The recurrence of trajectories to a neighborhood of a region in the phase space can
be used to analyze important properties of dynamical systems. When this tool is
applied to experimental or numerical generated data the limit of infinitely small
recurrence interval(the Poincar{\'e} limit) is never achieved. In this article, we
present a new effect that appears due to the finite size of the recurrence interval
and changes the exponential decay of the distribution of recurrence times. The
results are analyzed for the logistic and H{\'e}non maps but are expected to apply
to a large class of chaotic dynamical systems.}  

\spliter

\section{\label{sec.I} Introduction}

Since it was settled, the Poincar{\'e} recurrence theorem has been the source
of a number of paradoxes relating reversible microscopic dynamics of a system
on the one hand and irreversible macroscopic behavior of this system on the
other hand. An answer to these paradoxes was given by Boltzmann, who adopted
the law of big numbers ($N\rightarrow\infty$, where $N$ is the number of
degrees of freedom of the system under study) and the recognition that the
Poincar{\'e} Recurrence Time (PRT) to a highly improbable initial condition is too
large to be observed in times normally available. Boltzmann's point of view
was recently restated by Lebowitz \cite{lebowitz} in contrast with a new point
of view based on nonlinear dynamics. In this new approach, rather than the
limit $N\rightarrow\infty$, the central role is played by the sensitivity to 
initial conditions added to the idea that the Poincar{\'e} recurrence time do not
need to be very large to lie beyond the observable range
limit~\cite{aquino,ruelle}.   

Besides its fundamental importance for classical statistical
mechanics~\cite{frisch,kac2}, PRT statistics has been used, in the recent
years, as a tool for time series analysis in a variety of areas ranging from
economics to plasma physics
\cite{gao,murilo.bolsa,murilo.plasma1,murilo.plasma2}, and as a way of
studying trapping properties in Hamiltonian systems \cite{chirikov,meiss,hu},
which is an important feature for anomalous transport
processes~\cite{zas.primeiro,zaslavsky.rev}. The series of recurrence times
itself has also been subject of fractal
analysis~\cite{afraimovich,afraimovich2,hadyn}. All these aspects have brought
a renewed interest in the study of recurrence time statistics.  

General results have shown that the exponential-one law, $e^{-t}$, holds for
  the cumulative probability of first recurrence times (scaled by the mean first
  recurrence time) of transitive Markov chains~\cite{pitskel}, hyperbolic dynamical
  systems such as axiom A diffeomorphisms~\cite{hirata.axiomA} and for systems
  verifying a strong mixing property~\cite{hirata.phimixing}. Further improvement to
  the study of recurrences to finite size intervals has been given by Galves and Schmitt~\cite{galves}. They have
  computed an upper bound for the difference between the cumulative probability of
  first recurrence times and the exponential-one law. Moreover, they have shown that
  the right scaling of the first recurrence times should include an extra factor
  besides the mean first recurrence time. This factor lies between two strictly
  positive constants independent of the recurrence interval. Recently these results
  have been extended to unimodal maps~\cite{bruin}.

The Poissonian statistics, or its cumulative equivalent exponential-one law for
scaled recurrence times, is deduced in the limit of infinitely small recurrence
interval. However, in many of the recent applications mentioned previously the
recurrence times are obtained either from numerical simulations or experimental
data. In these cases it is unavoidable to use a recurrence interval with a finite
size. The size of the interval is chosen in order to obtain a sufficient number of
recurrences to build the recurrence time (RT) statistics. In this article we are
interested in the RT statistics to finite size intervals of chaotic dynamical
systems. 

We begin using simple basic concepts of combinatorial analysis to
deduce the statistics of recurrences
to a given finite size interval for random processes and chaotic systems with strong
mixing. This statistics applies not only for 
the first recurrence time (RT) but for all the $r$-th recurrence time. We
obtain this statistics, which we call binomial-like distribution of
RT, as a result of a simple combinatorial analysis problem. This distribution
is valid for every size of the return region. When the probability of coming
back to the return region is very small, the binomial-like RT statistics
reduces to the Poissonian statistics, commonly observed for PRT problems in
the literature~\cite{afraimovich2,murilo.plasma2,zaslavsky.pr}. Since we adopt a
combinatorial approach to deduce this statistics, almost no information about the
dynamics of the system, but its  strong chaotic mixing property, can be obtained
from it. Therefore, dynamical properties show their signature when the  recurrence
time statistics deviates from the former ones. One of these deviations is
particularly important for Hamiltonian systems and concerns with a power law tail
for long recurrence times \cite{zaslavsky.pr}. In this article, we study a type of
deviation which is related with the presence of unstable periodic orbits inside of
the recurrence interval.  We call it short time memory effect. This deviation
  originates the extra factor, which should multiply the mean recurrence time in
  order to give the right scaling of the series of recurrence times as considered by
  Galves and Schmitt~\cite{galves}. It appears when a finite recurrence interval
is considered. Although this deviation is restricted to a few short recurrence times,
it changes the whole statistics. Moreover, we would like to emphasize that our
deviation is in agreement with the bounds estimated in the previous works. 

The article is structured as follows: in section~\ref{sec.binomial}, we present
the deduction of the binomial-like statistics, which is exemplified by a 
Gaussian stochastic process. The statistics for chaotic dynamical systems
(logistic and H{\'e}non map) are calculated numerically in
section~\ref{sec.numerical}, where the short time memory effect is clearly
observed. In section~\ref{sec.memory}, we explore the origins of this effect and how
it changes the RT statistics. Finally, in section~\ref{sec.conclusion}, we summarize
our conclusions of this article. 

\section{\label{sec.binomial} Binomial-like Distribution}

Let $f:M\rightarrow M$ be a homeomorphism with an invariant measure
$\mu(M)=1$. Given a region $I\subset M$ with $\mu(I)>0$, the Poincar{\'e}
Recurrence Theorem asserts that a trajectory, having started inside $I$, returns
to $I$ infinitely many times. The time interval, $T_{i}$, between the $i$-th
and the $(i+r)$-th return is what we refer to as the $r$-th recurrence
time. This time interval is just one of an infinite sequence
$\{T_{i}:i=1,2,\ldots,\infty\}$, and we are interested in the statistics of
this sequence. 

\begin{figure}[!ht]
\centerline{
\includegraphics[width=\columnwidth]{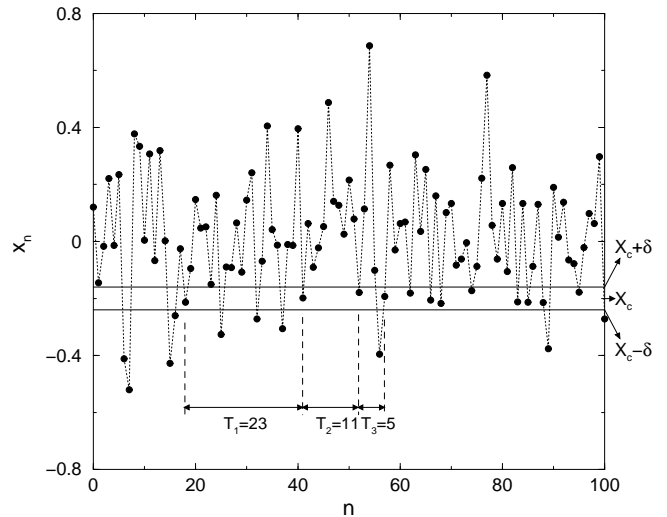}}
\caption{First recurrence time to the interval $I$ for a random time series
         with Gaussian distribution.}
\label{fig.serie}
\end{figure}

For convenience, most of the calculations are made for unidimensional
systems. In this case, the interval $I$ is defined as
$I(X_c,\delta)=[X_c-\delta,X_c+\delta]$, as illustrated in
FIG.~\ref{fig.serie} with a Gaussian random time series. When we have small
values of $\delta$, and thus a small probability $\mu(I)$, we are dealing with
the Poincar{\'e} recurrence time.
 
This article concerns the discrete time case, where the system is observed
at a constant sample rate $\tau=1$. A few adjustments are needed for the
continuous time case~\cite{kac,balakrishnan}.

In order to obtain the statistics for the recurrence time, consider the
following simple problem: Let $e_{1}$ and $e_{2}$ be two mutually exclusive
events. The event $e_{1}$ occurs with the constant probability $\mu$ and
$e_{2}$ with the constant probability $(1-\mu)$. Consider now a
sequence of $T_{i}$ trials $\{S_{k}:k=1,2,\ldots,T_{i}\}$ where $S_{k}=e_{1}$
or $S_{k}=e_{2}$. What is the probability of having $r$ events of type
$e_{1}$ and $(T_{i}-r)$ events of type $e_{2}$ with the constraint that the
last trial results is an event of the type $e_{1}$? This kind of problem is
known in the literature of combinatorial analysis as Bernoulli trials ({\em
  ``repeated independent trials for which there are only two possible outcomes
  with probabilities that remain the same throughout the trials''}
\cite{feller}).

The answer for this problem is the following: the probability of having
$r$ events $e_{1}$ and $(T_{i}-r)$ events $e_{2}$ is
$\mu^{r}(1-\mu)^{T_{i}-r}$. The last event must be of the type $e_{1}$,
then there are
\begin{equation*}
\label{eq.ways}
\dfrac{(T_{i}-1)!}{(T_{i}-r)!(r-1)!}
\end{equation*}
ways of having $(r-1)$ events $e_{1}$ in the previous $(T_{i}-1)$
trials. Combining these results and suppressing the index $i$, since $T_{i}$ is
just one of an infinite sequence of $r$-th recurrence times, we have: 
\begin{equation}
\label{eq.binomial.r}
P(T;r,\mu)=\dfrac{(T-1)!}{(T-r)!(r-1)!}\mu^{r}(1-\mu)^{T-r}\,.
\end{equation}

For a dynamical system with an invariant ergodic measure and for which each
step is independent from the previous ones, it is easy to see that
Eq.~(\ref{eq.binomial.r}) gives the probability of $r$-th recurrence time if
we consider the following analogy: between the $i$-th and the $(i+r)$-th
return to the interval $I(X_{c},\delta)$ the trajectory spends $T_{i}$ steps,
each step (one trial) has a probability $\mu=\mu(I(X_{c},\delta))$ of being in
the interval $I(X_{c},\delta)$ (event $e_{1}$) and a probability
$[1-\mu(I(X_{c},\delta))]$ of being outside $I(X_{c},\delta)$ (event $e_{2}$). 

Usually one is interested in the first recurrence time statistics ($r=1$). In
this case, Eq.~(\ref{eq.binomial.r}) gives 
\begin{equation}
\label{eq.binomial.1}
P(T;1,\mu)=\mu (1-\mu)^{T-1} \,,
\end{equation}
that can be rewritten as
\begin{equation*}
\label{eq.binomial.exp}
P(T;1,\mu)=\dfrac{\mu}{1-\mu}e^{\ln(1-\mu)T}\,,
\end{equation*}
which, by its turn, reduces to the Poissonian statistics
\begin{equation}\label{eq.poisson.1}
P(T;1,\mu)=\mu e^{-\mu T}\,,
\end{equation}
when $\mu\rightarrow 0$.

This statistics is the one commonly encountered for Poincar{\'e} recurrences in
chaotic dynamical systems \cite{gao,afraimovich2}. The small
$\mu(I(X_{c},\delta))$ condition is, usually, satisfied when we take small
values of $\delta$. 

For any $r$, in the usual limit~\cite{feller}, $\mu\rightarrow 0$ and
$T\gg 1$, the binomial-like distribution 
(Eq.~(\ref{eq.binomial.r})) reduces to the Poisson-like distribution
\begin{equation}\label{eq.poisson.r}
P(T;r,\mu)=\dfrac{\mu(\mu T)^{r-1}}{(r-1)!}e^{-\mu T}\,.
\end{equation}

Eqs.~(\ref{eq.binomial.1}) and (\ref{eq.poisson.1}), as well as any other
distribution of first RT, must satisfy two conditions: the first one
is, obviously, the normalization, 
\begin{equation}\label{eq.normalization}
\sum_{T=1}^{\infty} P(T;1,\mu)=1\,,
\end{equation}
and the second one is the Kac's lemma \cite{kac,kac2,cornfeld}:
\begin{equation}\label{eq.kac}
\langle T \rangle\equiv\sum_{T=1}^{\infty} T\,P(T;1,\mu)=\dfrac{\tau}{\mu}\,.
\end{equation}
 Although Kac's lemma is usually applied to closed Hamiltonian systems, its
  original  derivation \cite{kac} was based in general assumptions that cover a
  large class of dynamical systems including those we shall discuss in this work.

 The above conditions will be used in section~\ref{sec.memory} to take into
account dynamical effects on the recurrence time statistics.

\begin{figure}[!ht]
\centerline{
\includegraphics[width=\columnwidth]{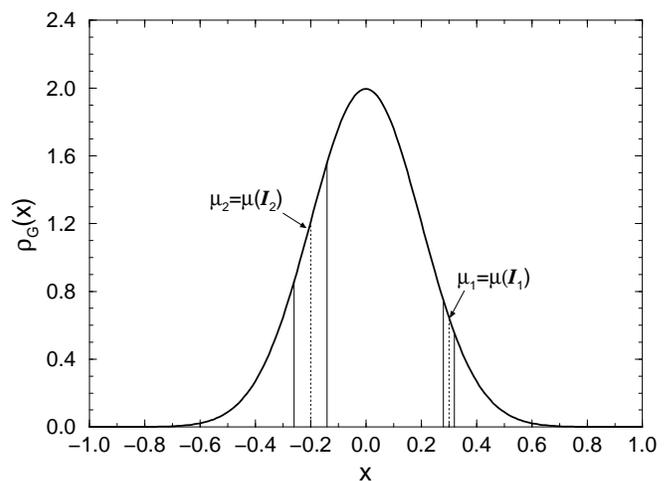}}
\caption{Gaussian density of probability with $\langle x \rangle=0$ and
  $\sigma=0.2$. The probabilities of returning to the regions $I_{1}$ and
  $I_{2}$ is given by $\mu_{1}$ and $\mu_{2}$, respectively.}
\label{fig.dist.gauss}
\end{figure}

To finalize this section we shall exemplify the binomial-like distribution
obtaining the RT statistics of a stochastic process whose variable, $x$, is
governed by a Gaussian density of probability: 

\begin{equation}
\label{eq.dist.gauss}
\rho_G(x)=\dfrac{1}{\sqrt{2\pi\sigma^2}} e^{-\dfrac{(x-\langle x
    \rangle)^{2}}{2\sigma^2}}\,, 
\end{equation}
with $\langle x \rangle =0$ and $\sigma=0.2$ (see FIG.~\ref{fig.dist.gauss}).
 
The probability, $\mu(I(X_{c},\delta))$, of returning to an interval
$I(X_{c},\delta)$ is given by:
\begin{equation}
\label{eq.mu}
\mu(I(X_{c},\delta))=\int_{X_{c}-\delta}^{X_{c}+\delta} \rho_{G}(x)dx\,,
\end{equation}
in particular, for the intervals $I_{1}$ and $I_2$, shown in
FIG.~\ref{fig.dist.gauss}, we have $(X_{c1}=0.3,\delta_{1}=0.02)$ and
$(X_{c2}=-0.2,\delta_{1}=0.06)$, respectively. Their corresponding
probabilities are $\mu(I_1)=0.02596$ and $\mu(I_2)=0.09679$.

\begin{figure}[!ht]
\centerline{
\includegraphics[width=\columnwidth]{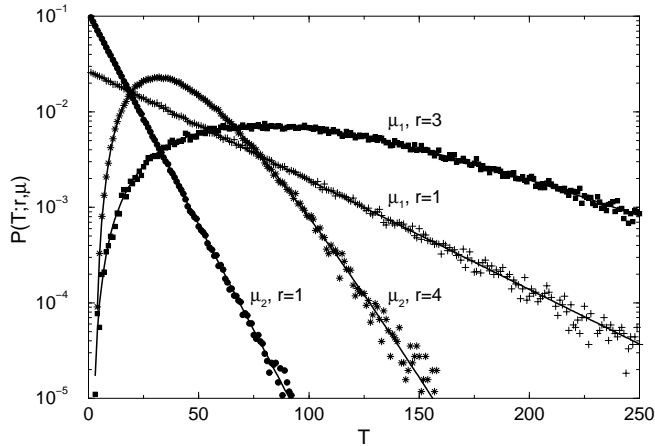}}
\caption{The $r$-th RT statistics for a Gaussian random walk. We chose
 arbitrary values of~$r$ and the two intervals illustrated in
  FIG.~\ref{fig.dist.gauss}. The solid lines are given by Eq.~(\ref{eq.binomial.r}).} 
\label{fig.ret.gauss}
\end{figure}

FIG.~\ref{fig.ret.gauss} shows the statistics of recurrences to the intervals
$I_{1}$ and $I_{2}$ for the Gaussian random time series.
The solid lines correspond to the analytical results given by
Eq.~(\ref{eq.binomial.r}) and they are in a good agreement with the numerical
results.

\section{\label{sec.numerical} Chaotic dynamical systems}

Let's see now what happens when we consider the statistics of first recurrence of
two deterministic dynamical systems.  

\begin{figure}[!ht]
\centerline{
\includegraphics[width=0.8\columnwidth]{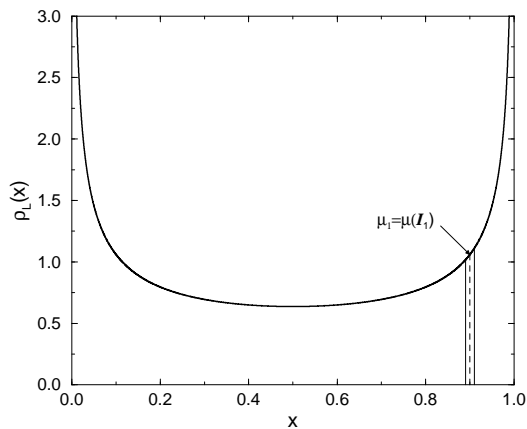}}
\caption{Invariant probability density for the logistic map. The marked area, 
  $\mu_{1}$, represents the measure of the interval
  $I_{1}(X_{c1}=0.9,\delta_{1}=0.01)$.} 
\label{fig.dist.log}
\end{figure}

The first system we analyze is the logistic map, 
\begin{equation}\label{eq.logistic}
x_{n+1}=b\,x_{n}(1-x_{n})\,,
\end{equation}
in the completely chaotic regime (parameter $b=4$). Its invariant probability
density,
\begin{equation}\label{eq.log.dist}
\rho_L(x)=\dfrac{1}{\pi \sqrt{x(1-x)}}\,,
\end{equation}
is shown in FIG.~\ref{fig.dist.log}. In this figure, we also see the return interval
$I_{1}$ ($X_{c1}=0.9,\delta_{1}=0.01$) with measure $\mu(I_1)=0.02124$.
 
\begin{figure}[!ht]
\centerline{
\includegraphics[width=\columnwidth]{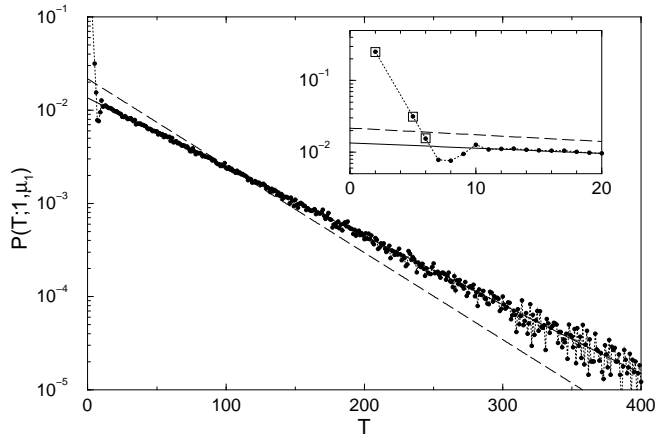}}
\caption{Recurrence time distribution for a logistic map ($I_1=[X_c=0.9,\delta=0.01]$
  and $N=10^{8}$ points) trajectory. The dashed line is the binomial-like
  distribution for $\mu(I_1)=0.02124$ (which coincides with the corresponding
  Poissonian distribution). The full line is the curve
  calculated to satisfy~(\ref{eq.normalization}) and~(\ref{eq.kac}) after
  $T=n^{*}=10$. The inset is an amplification of the small return times region and
  the squares are the analytic calculation of the return time probability as
  explained in FIG.~\ref{fig.t2.log}.} 
\label{fig.X0.9}
\end{figure}

Using the logistic map, we create a trajectory with $N=10^{8}$
points and look for recurrences to the interval
$I_{1}=[X_{c1}=0.9,\delta_{1}=0.01]$, indicated in FIG.~\ref{fig.dist.log}. The
statistics of first recurrence times to this interval is shown in
FIG.~\ref{fig.X0.9}. 

As our second example of dynamical system we choose the H{\'e}non map,  
\begin{equation}\label{eq.henon}
\begin{split}
x_{n+1}&=1-a\,x_{n}^{2}-y_{n}\,,\\
y_{n+1}&=b\,x_{n}\,.
\end{split}
\end{equation}
with the parameters $a=1.4$ and $b=0.3$.

\begin{figure}[!ht]
\centerline{
\includegraphics[width=\columnwidth]{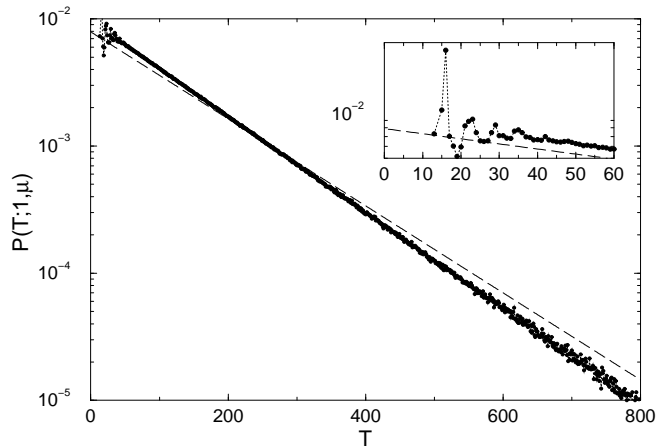}}
\caption{Recurrence time statistics for the H{\'e}non map with
  $I(\mathbf{X}_{c},\delta)$ given by $\mathbf{X}_{c}=(0.304, 0.210)$ and
  $\delta=10^{-2}$. The total of $10^{7}$ recurrence events were
  considered. Dashed line is the binomial-like distribution for
  $\mu(I)=\langle T \rangle^{-1}=0.00785$. The inset is an amplification
  of the short recurrence times region.}
\label{fig.henon}
\end{figure}

The distribution of first recurrence to a finite size interval for the H{\'e}non map
is shown in FIG.~\ref{fig.henon}. In order to obtain this result, we eliminate the
transient by iterating the H{\'e}non map $10^{3}$ times and take the final point,
$\mathbf{X}_{c}=(x_{c},y_{c})$, as the center of the phase space interval, $I$,
with radius $\delta=10^{-2}$. An initial condition inside the interval $I$ is chosen
and the H{\'e}non map is iterated until we obtain $10^{7}$ recurrence events.

Both maps present distributions of recurrence times which fall exponentially (a
straight line for a linear-log graphic) after some $T\geq n^{*}$ ($n^{*}\approx 10$
for the logistic map and $n^{*}\approx 40$ for the H{\'e}non map). What is indicated
through the solid lines in FIG.~\ref{fig.X0.9} and FIG.~\ref{fig.henon} for the
logistic map and H{\'e}non map, respectively. Nevertheless, these solid lines do not
coincide with the ones (dashed lines) given by the binomial-like statistics,
Eq.~(\ref{eq.binomial.1}), which is indistinguishable from what would be the correct
Poissonian statistics with $\langle T \rangle=\dfrac{1}{\mu(I)}$ for the particular
size of the intervals used. In particular, the difference between the right mean
recurrence time and the inverse of the slope of the solid line in
figure~\ref{fig.henon} is approximately $9.3\%$. This difference indicates that
  an extra factor, besides the mean recurrence time, should be considered in order
  to obtain the right scaling of the series of recurrence times~\cite{galves}. 

The reason for this deviation is the following: In the previous section, we showed
that the complete independence of iteration and the existence of an invariant
measure were necessary to obtain the binomial-like statistics of recurrence to
an interval of finite size. However, deterministic dynamical systems hardly
  fulfill the condition that an iteration is completely independent from the previous
ones. Therefore, dynamical systems have a kind of memory which affects the
distribution of short recurrence times as we shall show in the next section.  

\section{\label{sec.memory} Memory Effects on the RT Distribution}

In this section, we show how the memory effect changes the
probability of short recurrence times, and how the latests are responsible
for the deviation of the whole distribution. 

We take the advantage of working with a simple dynamical system, the logistic
map, which gives us the possibility of calculating analytically the probability of
short recurrence times.   

\subsection{\label{subsec.analytic} Short Recurrences and Lack of Memory}

\begin{figure}[!ht]
\centerline{
\includegraphics[width=0.8\columnwidth]{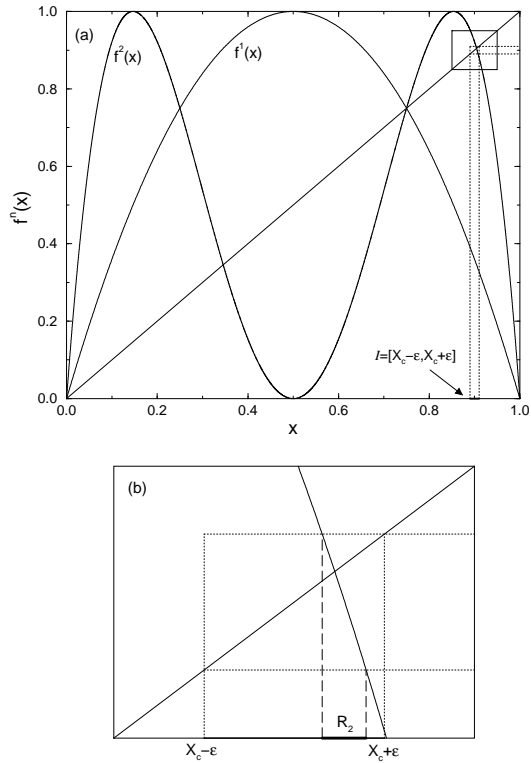}}
\caption{Analytic calculation for the return time probability for $T=2$ in the
  logistic map. The region $R_{2}$ represents the points inside $I$ that
  return to $I$ after two iterations of the map. We also see that $R_{1}=\varnothing$.}
\label{fig.t2.log}
\end{figure}

The procedure to obtain the probability of short recurrence times, illustrated in
FIG.~\ref{fig.t2.log}, consists in identifying the region, $R_{n}\subset I$, of
  initial points whose trajectories return to $I$ after $n$ iterations of the map
$f(x)$, what is equivalent to the first recurrence of the map $f^n(x)$. 
Since we are specially interested in the first return time, we
note that, if $R_{n} \cap R_{m}\neq\varnothing$ for $m<n$, the points from the
intersection will return to $I$ in $m$, not $n$, iterations. To avoid this
error, we define a new interval $\bar{R}_{n}=R_{n}-(R_{n}\cap R_{m})$ for every $
m<n$. The probability of a first recurrence time $T=n$ is, then, given by: 
\begin{equation}\label{eq.P1} 
P_{1}(T=n)=\dfrac{\mu(\bar{R}_{n})}{\mu(I)}\,.
\end{equation}

This method provides the probability of any first recurrence time.
Nevertheless, the determination of the regions $\bar{R}_{n}$ becomes a
cumbersome task as $n$ grows. 

Although the above procedure is not useful for calculating the probabilities of
first return when $n$ becomes large, it is useful for showing how the chaotic
dynamics simulates a random process for large recurrence times.  

When $n$ grows, the region $R_{n}$ becomes the union of a large number, $N$, of
disjoint sub-regions in the interval~$I$, that is:
\begin{equation}\label{eq.measure_R_n}
R_{n}=\bigcup_{i=1}^{N}R^{i}_{n}\,,
\quad \text{with} \quad
\mu(R_{n})=\sum_{i=1}^{N}\mu(R^{i}_{n})\,.
\end{equation}
The more sub-regions we have, the smaller they are. 

In particular, as the map $f^{n}(x)$ is a polynomial of order $2^{n}$ there
are $2^{n}$ unstable periodic orbits, each one with its associated
sub-region. The $2^{n}$ sub-regions  are distributed in the interval $[0,1]$ according
to the invariant density, $\rho_L(x)$, given by Eq.~(\ref{eq.log.dist}). The
sub-regions $R^{i}_{n}$ are just the ones which are inside the interval $I$. 

For large $n$, there are $N\approx 2^{n}\mu(I)$ sub-regions in the interval
$I$, each one with the same measure,
\begin{equation}\label{eq.measure_R_i}
\mu(R^{i}_{n})=\dfrac{\mu(I)}{2^{n}}\,.
\end{equation}

The probability of having a recurrence to $I$ in the time $T=n$ is:
\begin{equation}\label{eq.P}
P(T=n)=\dfrac{\mu(R_{n})}{\mu(I)}\,.
\end{equation}

Putting Eqs.~(\ref{eq.measure_R_n}) and (\ref{eq.measure_R_i}) in
Eq.~(\ref{eq.P}), we obtain the recurrence probability,
$P(T=n)\approx\mu(I)$, independent of $n$ for large $n$, which is the same
hypothesis assumed for the Bernoulli trials problem in
section~\ref{sec.binomial}. This argument justifies the exponential decay of
the recurrence time statistics after the short times.
It must be emphasized, however, that $P(T)$ given by Eq.~(\ref{eq.P}) is not the
probability of first recurrence to $I$ in the time $T=n$. This one is
given by Eq.~(\ref{eq.P1}), where $\bar{R}_{n}$, instead of $R_{n}$, is
used. It is in the calculations of $\bar{R}_n$ that the memory effect appears,
since the approximation $N\approx 2^n \mu(I)$ is not valid for $n<n^{*}$.
The same arguments hold for a general hyperbolic chaotic system, since the
number of recurrence sub-regions is equal to the number of periodic orbits, and
the latest grows as $e^{h T}$, where $h$ is the topological entropy.   

\subsection{The Fitting of the Memoryless Exponential}

Let's see how the deviation of the shortest recurrence times affects
the whole distribution. As it was shown in FIG.~\ref{fig.X0.9}, for greater values
of $T$ (after the decay of the memory, $T>n^{*}$) the recurrence time
distribution approaches a straight line in a linear-log graphics, what can be
generally represented by an exponential,   
\begin{equation}\label{eq.exp}
P_{exp}(T) =g_{0} e^{-\gamma T}.
\end{equation}
Since we know the mean recurrence time (or the measure of the recurrence interval),
the above two parameters, $g_{0}$ and $\gamma$, can be analytically obtained by
using the two conditions presented in section~\ref{sec.binomial}, namely, the
normalization:  
\begin{equation}\label{eq.fitting.1}
\sum_{T=1}^{n^{*}-1} P_{1}(T)+\sum_{T=n^{*}}^{\infty} P_{exp}(T)=1\,,
\end{equation}
and the Kac's lemma:
\begin{equation}\label{eq.fitting.2}
\langle T \rangle=\sum_{T=1}^{n^{*}-1} TP_{1}(T)+\sum_{T=n^{*}}^{\infty}
TP_{exp}(T)=\dfrac{1}{\mu(I)}\,,
\end{equation}
where $P_{1}(T)$ is given by Eq.~(\ref{eq.P1}) or obtained directly from the
data. The agreement of the exponential (Eq.~(\ref{eq.exp})) obtained by this
procedure with the linear part
(in the linear-log representation) of the RT distribution for the
logistic map, as shown by the solid lines in FIG.~\ref{fig.X0.9} and
FIG.~\ref{fig.henon}, was verified for different recurrence intervals.  

\subsection{\label{ssec.dependece} Dependence of the memory effect on the interval}

In this sub-section, we explore the dependence of the memory effect on
the position~$X_c$ and size of the interval with a special interest in small
intervals. As we argued before, only a few 
number of short recurrence times diverge from a straight line  
in the linear-log representation, but their effects on the whole distribution is
considerable. So we can take the deviation of the asymptotic exponential
from the binomial (Poissonian) distribution as a
quantifier of the short time memory effect. Let 
\begin{equation*}\label{eq.fit}
g(T)=g_{0}e^{-\gamma T}\,,
\end{equation*}
be the exponential adjusted to the asymptotic part of the distribution. As
shown in section~\ref{sec.binomial}, the
binomial distribution results in~$\gamma=-ln(1-\mu)$, that reduces
to~$\gamma=\mu$ in the Poissonian case. Since we are specially interested
in small intervals, we will use the relative deviation from the Poisson
statistics~$\gamma/\mu(I) - 1 = \gamma \langle T \rangle -1$ as the quantifier
of the memory effect.  

\begin{figure}[!ht]
\centerline{
\includegraphics[width=\columnwidth]{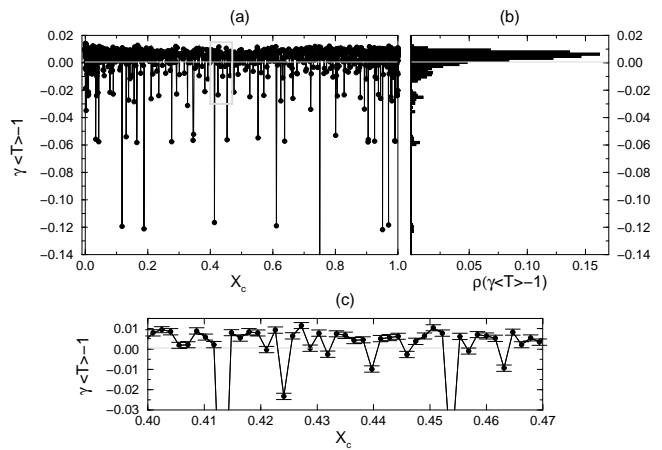}}
\caption{Dependence of the  memory effect~$(\gamma \langle T \rangle -1)$ on
  the position of the recurrence interval~$(X_c)$. (a) Results for the $10^3$
  intervals with measure~$\mu(I)=10^{-3}$, centered in~$X_c$, that covers the
  interval~$x \in [0,1]$ of the logistic map. (b) histogram of (a). (c)
  amplification of (a) showing the size of the fitting errors. The gray line
  represents the expected value for the binomial distribution.}
\label{fig.Xc}
\end{figure}

In FIG.~\ref{fig.Xc}, it is shown the dependence of the memory effect with the
position of the interval. We note that the effect occurs for virtually all
recurrence intervals. Most of the
intervals deviate from the exponential positively but for recurrence intervals
that contain periodic orbits of small periods the asymptotic exponential
is negatively deviated when compared to the binomial distribution. 

\begin{figure}[!ht]
\centerline{
\includegraphics[width=\columnwidth]{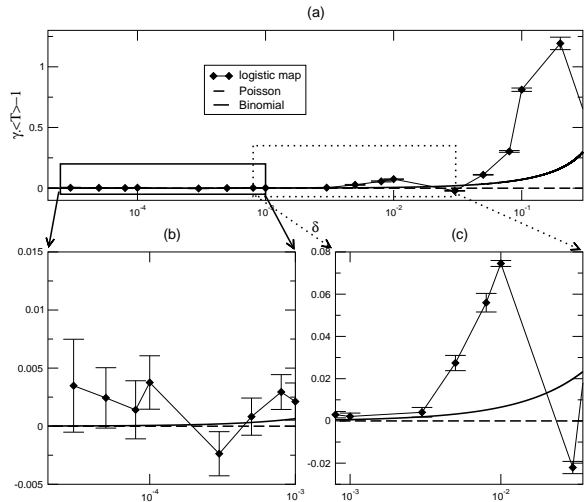}}
\caption{Dependence of the memory effect~$(\gamma \langle T \rangle -1)$ on
  the size of the recurrence interval~$(\delta)$. The position of the interval
  was chosen randomly as~$X_c=0.582$. The dashed and solid lines are the
  expected results for the Poisson and binomial distribution
  respectively. (b) and (c) are amplifications of (a) 
  for small values of~$\delta$ and show the convergence to the binomial
  statistics for~$\delta < 10^{-4}$ in the fitting precision.}
\label{fig.delta}
\end{figure}

To explore the dependence of the memory effect on the size of the interval we
took, by chance, an arbitrary position~$X_c=0.582$ and we vary its
semi-width, $\delta$. The results obtained numerically are illustrated in
FIG.~\ref{fig.delta}. For the logistic map, the relation between the 
measure, $\mu(I)$, and the semi-width, $\delta$, is easily found out
integrating the distribution~(\ref{eq.log.dist})
\begin{equation}\label{eq.mudelta}
\begin{split}
\mu(X_{c},\delta)&=\int_{X_{c}-\delta}^{X_{c}+\delta} 
\dfrac{1}{\pi\sqrt{(x(1-x))}}\,dx\\
&=\arcsin(X_{c}+\delta)-\arcsin(X_{c}-\delta)\,.
\end{split}
\end{equation}

With the relation~(\ref{eq.mudelta}) and remembering that for the binomial
distribution~$\gamma=-ln(1-\mu)$ we obtain the solid line in
FIG.~\ref{fig.delta}. From FIG.~\ref{fig.delta}, we see that for great values
of~$\delta$ the results deviate from both the Poissonian (dashed line) and
binomial-like distributions. The convergence of the numerical results of the
logistic map to the Poissonian statistics occurs in the limit of small interval,
coherently with  what is found in the literature~\cite{bruin}. This convergence occurs
clearly only for~$\delta < 10^{-4}$. Considering the error bars obtained by the
numerical fitting, these are much smaller intervals than the ones for which the
convergence of the binomial-like to Poisson takes place. Considering our previous
discussions, the convergence of the numerical results to the Poissonian statistics
occurs only when the short time memory effect becomes negligible.   

\section{\label{sec.conclusion}Conclusions}
Based on a simple Bernoulli trials problem, we obtain a binomial-like
distribution for the $r$-th recurrence time statistics of a generic interval.
This distribution depends only on the measure of the recurrence interval
($\mu(I)$) that, when  it is sufficiently small, turns out the statistic to
the usual Poissonian distribution for the first Poincar{\'e} recurrence time. 

The information related to the dynamical properties
appears in the deviation from these distributions as, for example, the
power-law behavior for large recurrence times studied in
references~\cite{chirikov,zas.primeiro}. In this article, we discuss a kind of
deviation that appears for finite size intervals. In this case the short recurrence
times are affected by a kind of memory of the chaotic systems. We show that the
origin of this short time memory effect is due to the existence of unstable periodic
orbits inside of the finite size recurrence interval. The analytical method of
calculating the recurrence probabilities, described in
section~\ref{subsec.analytic}, shows how these periodic orbits changes the
distribution of short recurrence times. Furthermore, our observed deviations for the
specific systems studied in this work are in agreement with bounds predicted for
general classes of systems in previous
works~\cite{pitskel,hirata.axiomA,hirata.phimixing,galves}.

The exponential growth of the number of periodic unstable orbits in chaotic dynamical
systems restates the condition of independence between recurrences for large times,
resulting in an exponential-like behavior of the recurrence time 
distribution after the decay of memory. Imposing the normalization condition and
the Kac's lemma, we are able to make an analytical fitting to the memoryless part
of the distribution. This fitting illustrates how the short recurrence times,
which are affected by the memory, modify the whole distribution. Since just a few
points deviate from an exponential, a histogram bin larger than one may lead to the
wrong conclusion that the distribution is Poissonian. What results in a
contradiction: the inverse of the false Poissonian  exponent is different from the
mean recurrence time obtained from the recurrence series. For example, for the data
of FIG.~\ref{fig.henon} this difference is of $9.3\%$. We believe that this alert
has to be taken into account when the RT statistics is calculated. We would like
to stress that our results indicate that the correct normalization of a series of
recurrence times should consider an extra factor, besides the mean recurrence time,
in order to properly obtain the asymptotic exponential-one law.  

We emphasize that the memory effect, discussed in this article, applies for any
recurrence interval of a general chaotic dynamical system. For small intervals it
may not be relevant because it turns smaller than the numeric precision. When one is
calculating RT numerically, it must be checked that the interval is sufficiently
small that this regime is already achieved. 


\begin{acknowledgments}
This work was made possible through financial support from the following
Brazilian research agencies: FAPESP and CNPq.
\end{acknowledgments}


\end{document}